# Mid-IR frequency measurement using an optical frequency comb and a long-distance remote frequency reference

B. Chanteau, O. Lopez, F. Auguste, B. Darquié, C. Chardonnet, A. Amy-Klein

Laboratoire de Physique des Lasers (LPL)
CNRS and Université Paris 13
Villetaneuse, France
bruno.chanteau@univ-paris13.fr

W. Zhang, G. Santarelli, Y. Le Coq

LNE-SYRTE
Observatoire de Paris, CNRS, UPMC
Paris, France

*Abstract*—We have built a frequency chain which enables to measure the absolute frequency of a laser emitting in the 28-31 THz frequency range and stabilized onto a molecular absorption line. The set-up uses an optical frequency comb and an ultrastable 1.55 µm frequency reference signal, transferred from LNE-SYRTE to LPL through an optical link. We are now progressing towards the stabilization of the mid-IR laser via the frequency comb and the extension of this technique to quantum cascade lasers. Such a development is very challenging for ultrahigh resolution molecular spectroscopy and fundamental tests of physics with molecules.

## I. Introduction

Ultrahigh resolution molecular spectroscopy is an alternative tool to atoms to perform tests of fundamental physics, such as the search for the non conservation of parity [1], the temporal stability of the electron-to-proton mass ratio [2], or the electron electric dipole moment [3]. Many of these tests rely on the availability of ultrastable and accurate laser sources emitting in the mid-IR where molecules exhibit rovibrational transitions. It is thus very challenging to develop a frequency stabilization or measurement scheme in the mid-IR which does not depend on quite rare frequency secondary references. Such high sensitivity spectroscopy experiments have been performed using an optical frequency comb. In that case, the IR frequency is compared to a high-harmonic of the comb repetition rate with a sum or difference frequency generation in a non linear crystal and a Ti:Sa optical frequency comb. It has been demonstrated at 10 µm with a $CO_2$ laser [4], at 3.4 µm using a HeNe laser [5], and recently extended to quantum cascade lasers at 5 µm [6], and 3.4 µm using a fiber comb [7].

## II. Experimental set-up

In our case, we have built a frequency chain with two fiber femtosecond lasers, a remote optical reference and an optical link, which enables to measure the absolute frequency of a $CO_2$ laser, emitting in the 28-31 THz frequency range, against an optical reference from the LNE-SYRTE (Fig. 1). The $CO_2$ laser, located at LPL, is stabilized onto a molecular absorption line.

The optical reference is a fiber laser, emitting at 1.55 µm, locked to a very high finesse cavity with the Pound-Drever-Hall method. A fiber femtosecond laser, centered at 1.55 µm, is phase-locked to the ultrastable laser. The repetition rate of this femtosecond laser is then compared to the primary standards of LNE-SYRTE (H-maser, cryo-oscillator, caesium fountain). This scheme enables to measure the absolute frequency of the ultrastable laser and therefore compensate the cavity drift, which is critical for some high sensitive spectroscopy experiments, as for example the parity violation experiment [1]. This optical reference is then sent through a 43-km long optical link to LPL [8].

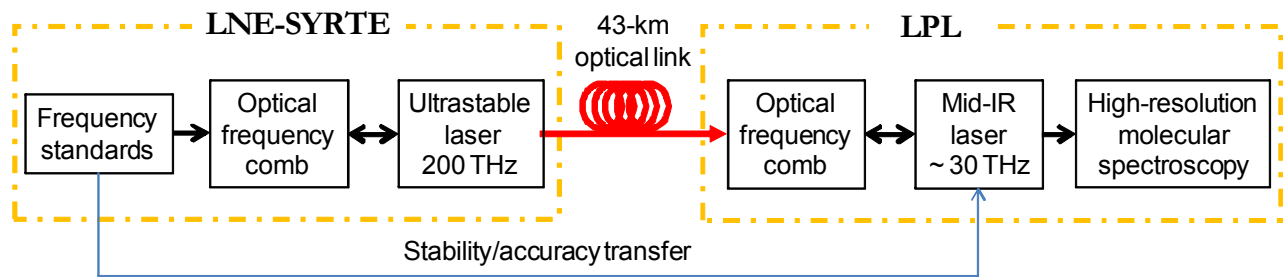

Figure 1. Mid-IR frequency measurement principle

The authors acknowledge funding support from *ANR*, *LNE*, *CNRS* and *Université Paris 13*.

At LPL, we aim at measuring ?? the frequency of a CO$_2$ laser, at around 10 µm, locked to an absorption line of the molecule OsO$_4$. A fiber femtosecond laser at 1.55 µm is used to transfer the optical reference stability and accuracy to the IR laser. The modes frequencies can be written as:

$$\nu_p = p f_{rep} + f_{CEO} \quad (1)$$

where $p$ is an integer near 780 000. The optical reference is used to lock the repetition rate of the comb. After removing the offset frequency $f_{CEO}$, we have:

$$f_{rep} = \frac{1}{N}(\nu_{ref} - \Delta_{ref}) \quad (2)$$

with $f_{rep}$ the repetition rate, $N$ an integer near 780 000, $\nu_{ref}$ the optical reference frequency, and $\Delta_{ref}$ the beatnote between the ultrastable laser and the femtosecond laser. Then, low-frequency modes are generated in a non linear fiber in order to obtain a comb centered around 1.85 µm. This comb is mixed with the CO$_2$ laser in a non linear crystal leading, via sum frequency generation, to a shifted comb of frequencies:

$$\nu_q + \nu_{CO_2} = q f_{rep} + f_{CEO} + \nu_{CO_2} \quad (3)$$

with $\nu_q$ the frequency of one mode of the comb, $\nu_{CO_2}$ the CO$_2$ laser frequency, $q$ an integer around 660 000. The beatnote $\Delta_{CO_2}$ between this shifted comb and the comb directly issued from the femtosecond laser, with modes $\nu_p$ gives:

$$\Delta_{CO_2} = (q f_{rep} + f_{CEO} + \nu_{CO_2}) - (p f_{rep} + f_{CEO})$$
$$= \nu_{CO_2} - (p-q) f_{rep} \quad (4)$$

In combination with (2), we obtain:

$$\nu_{CO_2} = \frac{p-q}{N}(\nu_{ref} - \Delta_{ref}) - \Delta_{CO_2} \quad (5)$$

with $p-q$ is near 120 000. Thus the CO$_2$ laser frequency is compared to a high-harmonic of the comb repetition rate [4]. Finally, the CO$_2$ laser frequency can be measured against the frequency standards of LNE-SYRTE.

## III. RESULTS AND PERSPECTIVES

With this set-up, we measured the frequency of a CO$_2$ laser, locked to an OsO$_4$ saturation line, with a relative stability of $6 \times 10^{-14}$ after 1 s integration time, not limited by the measurement set-up [9], even without compensating the optical link used for the experiments (Fig. 2). The measurements duration is however limited by the chemical stability of OsO$_4$ molecules, which degrade after a few hours. We have also measured the absolute frequency of the CO$_2$ laser, and we obtained a value consistent with the previous measurements [10]-[11].

We are now progressing towards the stabilization of the CO$_2$ laser via the frequency comb. Because molecules degrade over a few hours, the CO$_2$ laser frequency drifts. First results of stabilization show that this drift has been removed. The next step is to lock the CO$_2$ laser directly to the optical reference, without prestabilization on a molecular absorption line. This is necessary for the parity non conservation experiment since left- and right-handed molecules lines are alternately measured in a supersonic beam [1].

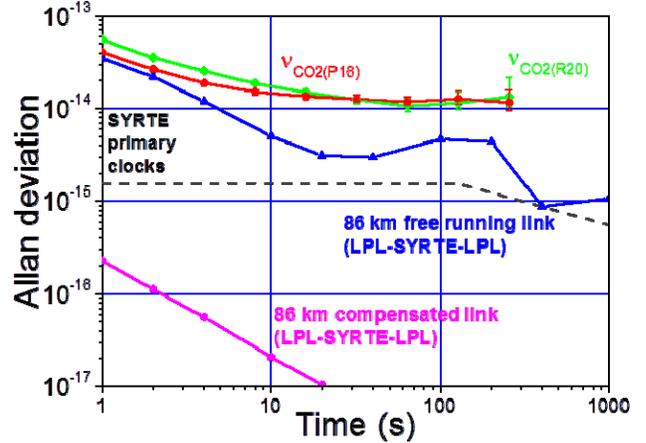

Figure 2. Allan deviation of the CO$_2$ laser frequency locked to an absorption line of OsO$_4$. In red and green, this is respectively the P(16) CO$_2$ laser line near the $^{190}$OsO$_4$ P(55) line and the R(20) CO$_2$ laser line near the $^{192}$OsO$_4$ R(67) line. In blue and magenta, typical stability of optical link used for the IR frequency measurement experiments. Dashed line : combination of the ultrastable laser and primary standards of LNE-SYRTE stabilities.

In the future, we will extend this stabilization/measurement scheme to quantum cascade lasers. Among several constraints, the ideal molecule for the parity violation experiment must have a rovibrational absorption line in the spectral range of an ultrastable laser. Till now, we were limited to the CO$_2$ laser, which emits in spectral windows about 1 GHz wide every 30-50 GHz, between 9 and 11 µm. With a quantum cascade laser, we can access a continuous spectral range of 200 GHz, at any wavelength between 2 and 20 µm. This would simplify the optical scheme of any ultrahigh sensitivity IR spectroscopy measurement.